%% file: main.tex
\documentclass[runningheads]{llncs}
\usepackage{graphicx}
\usepackage{makeidx}
\usepackage{url}
\usepackage{multirow}
\usepackage{cite}
\usepackage{graphicx}
\graphicspath{{figs/}}
\usepackage{amsmath,bm}
\usepackage{amssymb}
\usepackage{algorithm}
\usepackage{algorithmic}
\begin{document}
\title{How can spherical CNNs benefit ML-based diffusion MRI parameter estimation?\thanks{This work is supported by the EPSRC-funded UCL Centre for Doctoral Training
in Medical Imaging (EP/L016478/1), the Department of Health’s
NIHR-funded Biomedical Research Centre at UCLH and the Wellcome Trust.}}
\titlerunning{How S-CNNs benefit ML-based dMRI parameter estimation?}
\author{Tobias Goodwin-Allcock\inst{1} \and
Jason McEwen\inst{2} \and
Robert Gray\inst{3} \and
Parashkev Nachev\inst{3} \and
Hui Zhang\inst{1}}
\authorrunning{T. Goodwin-Allcock et al.}
\institute{Department of Computer Science and Centre for Medical Image Computing, University College London, London, United Kingdom \and
Kagenova Limited, Guildford, United Kingdom \and Department of Brain Repair \& Rehabilitation, Institute of Neurology, UCL, London, United Kingdom.\\
}
\maketitle              % typeset the header of the contribution
\begin{abstract}
This paper demonstrates spherical convolutional neural networks (S-CNN) offer distinct advantages over conventional fully-connected networks (FCN) at estimating scalar parameters of tissue microstructure from diffusion MRI (dMRI). Such microstructure parameters are valuable for identifying pathology and quantifying its extent. However, current clinical practice commonly acquires dMRI data consisting of only 6 diffusion weighted images (DWIs), limiting the accuracy and precision of estimated microstructure indices.
Machine learning (ML) has been proposed to address this challenge. However, existing ML-based methods are not robust to differing gradient schemes, nor are they rotation equivariant. Lack of robustness to gradient schemes requires a new network to be trained for each scheme, complicating the analysis of data from multiple sources. A possible consequence of the lack of rotational equivariance is that the training dataset must contain a diverse range of microstucture orientations. Here, we show spherical CNNs represent a compelling alternative that is robust to new gradient schemes as well as offering rotational equivariance. We show the latter can be leveraged to decrease the number of training datapoints required.

\end{abstract}

\input{00introduction}
\input{01networksBackground}
\input{02method}
\input{03resultsNdiscussion}

\input{04conclusion}

\bibliographystyle{splncs04}
\bibliography{references}

\end{document}

%% file: 00introduction.tex
\section{Introduction}
\label{sec:introduction}

Diffusion MRI (dMRI) plays an important role in neuroscientific  and clinical research because it can help infer tissue microstructure \cite{Bodini2014DiffusionDisorders}. 
To infer tissue microstructure from dMRI, we use mathematical models to estimate parameters from dMRI data.
At each voxel, the measurements made according to some acquisition scheme – commonly consisting of gradient schemes coupled with their diffusion sensitising factors – are fitted to mathematical models, such as the diffusion tensor (DT) \cite{Basser1994EstimationEcho}. 
From these models, dMRI parameters can be derived to reveal the microstructure such as fractional anisotropy (FA), which characterises the anisotropy of the tissue. 
The computation of these parameters from dMRI data is known as dMRI parameter estimation, which has traditionally been achieved with model fitting. 
However, the fidelity of dMRI parameters estimated in this way is limited by relatively high noise in the data, requiring more measurements to be acquired than what are routinely made in the clinic \cite{Jones2013GaussianSignal}.

As in many other fields, dMRI parameter estimation has recently been revolutionised by exploiting deep learning (DL), yielding greatly increased accuracy than conventional fitting when the acquisition scheme has a small number of measurements \cite{Golkov2016Q-SpaceScans,Aliotta2019HighlyNetworks}. 
However, current deep-learning methods, e.g. fully-connected networks (FCN), are ignorant of the acquisition scheme of a given acquisition, rendering these methods potentially not generalisable to new acquisition schemes. 
This complicates the application of a DL model to data acquired from multiple sources. 
Moreover, these methods do not exhibit rotational equivariance, a property that may help reduce the demand for training data. 

There have been a number of attempts to include the relationship between an acquisition scheme and the corresponding data\cite{Chen2020EstimatingNetworks, Park2021DIFFnet:B-values}. 
However, they do not utilise the topological features of the associated gradient schemes.
Gradient schemes for a given diffusion sensitising factor can be represented by points on the unit sphere.
Therefore, spherical convolutional neural networks (S-CNN), recently proposed as an alternative to FCNs \cite{Sedlar2021ADMRI, Elaldi2021EquivariantData},  provide a more natural solution to this problem.
However, currently there exists no direct evidence of the theoretical benefits of S-CNNs, such as rotational equivariance and robustness to different gradient schemes.
Here we aim to provide the very first empirical evidence of these advantages in the context of estimating rotation-invariant dMRI parameters. 

The rest of the paper is described as follows: Section \ref{sec:dMRIparamEstim} describes the dMRI parameter estimation problem and the conventional solution; Section \ref{sec:MLdMRIparamEstim} how machine learning attempts to solve this problem and the theoretical beneficial properties of S-CNNs; Section \ref{sec:method} then goes on to explain how we empirically test these properties; Section \ref{sec:resultsNdiscussion} summarises the results and discusses future work and Section \ref{sec:conclusion} provides the conclusion.

%% file: 01networksBackground.tex
\section{The dMRI parameter estimation problem and its conventional solution}
\label{sec:dMRIparamEstim}
This paper depends on an understanding of the dMRI parameter estimation problem. This problem, and its conventional solution, are explained in this section. As an example, we show how the dMRI parameter FA is estimated from a common clinical diffusion MRI acquisition consisting of 6 diffusion weighted images (DWIs).

\subsection{dMRI parameter estimation problem}
dMRI parameters can be used to infer or estimate properties of tissue microstructure. Although dMRI parameters typically cannot be directly measured, they modulate how the measured dMRI signals vary as a function of the acquisition setting; this function is known as the forward model. The dMRI parameters of interest are often the forward model parameters themselves, e.g. neurite density index in the NODDI model \cite{Zhang2012NODDI:Brain}, but can also be analytical functions of them, e.g. FA derived from the DT. The dMRI parameter estimation problem is about estimating these parameters from the acquisition scheme and the resultant measured signals.

\subsection{Conventional solution}
Conventionally, estimation of dMRI parameters is achieved by fitting the corresponding forward model to the measured dMRI signals. Let $\textbf{t}$ denote the dMRI parameters of the forward model and $\textbf{k}_i$ the acquisition setting, then the forward model  predicts the noise-free signal $S(\textbf{k}_i;\textbf{t})$. The fitting procedure finds  $\hat{\textbf{t}}$ the dMRI parameters that minimise the discrepancy between the noise-free signals predicted by the forward model and the measured signals as described in the following equation:
\begin{equation}
    \hat{\textbf{t}} = \arg\min_{\textbf{t}}(\sum_{i=1}^N{||s_i – S(\textbf{k}_i ; \textbf{t})||^2)},
\end{equation}
where the $N$ number of signals, $\textbf{s}=(s_1,...,s_N)$, are measured according to the acquisition scheme $\textbf{K}=(\textbf{k}_1,...,\textbf{k}_N)$.
However, because dMRI signals are inherently noisy, estimation from only a small number of measurements typically lead to noisy estimates of the dMRI parameters.

\subsection{FA estimation as an example}
\label{sec:dMRIparamEstim:subsec:DTexample}
FA is conventionally estimated using the diffusion tensor model, which is a common forward model of water diffusion in biological tissue on the micron scale. In the DT model \cite{Basser1994EstimationEcho}, tissue properties are modelled by a 3$\times$3 symmetric positive-definite matrix $D$, known as the diffusion tensor; due to its symmetry $D$ consists of 6 independent parameters.
The DT signal model is expressed here: 
\begin{equation}
S(b_i,\textbf{g}_i;D,S_0)=S_0 e^{-b_i \textbf{g}_i^T D \textbf{g}_i},   
\label{eqn:DTsignalmodel}
\end{equation}
using some typical diffusion MRI acquisition scheme $\textbf{K}=(\textbf{B},\textbf{G})$ where $\textbf{B}=(b_1,...,b_N)$ denotes the diffusion sensitising factors, also known as b-values - and $\textbf{G}=(\textbf{g}_1,...,\textbf{g}_N)$ the gradient scheme. The noise free signal in absence of diffusion gradients is $S_0$. Typical acquisitions for estimating DT include at least one b=0 measurement and at least 6 linearly independent non-b=0 measurements. This estimated diffusion tensor, $\hat{D}$, is then eigendecomposed and the resultant eigenvalues are used to calculate FA \cite{PeterJ.Basser1996MicrostructuralMRI}. 

For the common clinical acquisition we mentioned earlier, which consists of 6 dMRI signals and 1 b=0 signal, estimating FA with conventional fitting is much more straight forward but inaccurate. Under this condition the best estimate for $S_0$ is the b=0 signal. 
This allows us to simplify Eqn (\ref{eqn:DTsignalmodel}) by normalising the six dMRI signals. We are left with 6 normalised dMRI signals to estimate the DT's 6 independent parameters. Therefore, there is a direct 1-to-1 mapping between the normalised noisy dMRI signals and the estimated DT. In conventional fitting, from this estimated DT the FA is calculated. The inherent noise in dMRI signal causes poor DT estimation and, consequently, poor FA estimation. Therefore, a new method is required to estimate FA with greater accuracy.

The 1-to-1 mapping is a property that plays an important role in Section \ref{subsec:methods/GenDenseSampledSphSignals}. We explore this property later to generate an S-CNN compatible input. 

\section{ML solutions to the dMRI parameter estimation problem and the theoretical benefits of S-CNNs}
\label{sec:MLdMRIparamEstim}
Deep learning (DL) has been proposed as a solution to dMRI parameter estimation from small numbers of diffusion weighted images (DWI). This section provides (1) an example of the current machine learning standard for voxel-wise estimation (2) the theoretical limitations of this architecture (3) the architecture features that theoretically benefit S-CNNs. As an example, we show how the dMRI parameter FA is estimated from a common clinical diffusion MRI acquisition consisting of 6 DWIs.

\subsection{Deep learning solution}
Deep learning models, $F$, map the dMRI signals $\textbf{s}$ and their corresponding acquisition scheme - consisting of diffusion sensitising factors $\textbf{B}$ and gradient scheme $\textbf{G}$ -  directly to dMRI parameters $\textbf{t}$: 
\begin{equation}
    \textbf{t} = F(\textbf{s},\textbf{B},\textbf{G};\theta).
\end{equation}
This function is learnt by optimising the trainable parameters, $\theta$, on training data following:
\begin{equation}
    \hat{\theta} = \arg \min_{\theta} ( \sum_{j=1}^M{||\textbf{t}^{(j)}-F(\textbf{s}^{(j)},\textbf{B}^{(j)},\textbf{G}^{(j)};\theta)||^2)},
\end{equation}
where the $j^\text{th}$ example from $M$ training examples has signal input $\textbf{s}^{(j)}$, diffusion sensitising factors $\textbf{B}^{(j)}$ and gradient scheme $\textbf{G}^{(j)}$ with corresponding ground truth output $\textbf{t}^{(j)}$. 
Unlike conventional parameter estimation, inference for an unseen set of DWIs is performed without further optimisation as in the following equation:
\begin{equation}
    \hat{\textbf{t}} = F(\textbf{s},\textbf{B},\textbf{G};\hat{\theta}),
\end{equation}
where the estimated $\textbf{t}$ is $\hat{\textbf{t}}$.
The quality of this estimation depends on many factors. The two factors explored in this work are the training data distribution and the choice of network architecture.

\subsection{Fully-connected networks}
The first and most common deep learning network architecture applied to dMRI data is the FCN \cite{Golkov2016Q-SpaceScans, Aliotta2019HighlyNetworks}. Conventionally these have been implemented following:
\begin{equation}
t = F_\text{FCN}(\textbf{s};\theta_{FCN}), 
\end{equation}  
where $F_{FCN}$ is a fully-connected network with trainable parameters $\theta_{FCN}$. 
The network's input consists of the dMRI signals. 
Absent from the equation is the acquisition scheme so the network is ignorant of the acquisition scheme. 
Estimation from a new set of DWIs is accurate only if the acquisition scheme for the new data is consistent with the acquisition scheme during training \cite{Park2021DIFFnet:B-values}.

An FCN's architecture is not designed to be rotationally equivariant. A theoretical consequence of lacking rotational equivariance is that the training dataset may have to contain a diverse set of tissue microstructure orientations for FCNs to accurately estimate independent of fibre orientation.

\subsection{Spherical CNNs}
S-CNNs theoretically improve over FCNs - both in terms of robustness to the gradient scheme and robustness to the training data distribution - because of the difference in network architecture.

An S-CNN's architecture differs greatly to an FCN's but not in the way one may expect. In S-CNNs, the convolution isn't across multiple voxels, like traditional CNNs, but over the spherical image space. Therefore, S-CNNs are voxel wise networks just like FCNs. 
The spherical image is generated at each voxel from the dMRI signals, $\textbf{s}$, along with their corresponding gradient scheme $\textbf{G}$.
We see that this architecture may naturally address the highlighted limitations of FCNs.
Firstly, an S-CNN's input is informed of the gradient scheme as shown in the following equation:
\begin{equation}
    t = F_\text{S-CNN}(\textbf{s},\textbf{G};\theta_{\text{S-CNN}}),
\end{equation} where $F_{\text{S-CNN}}$ is an S-CNN with trainable parameters $\theta_{\text{S-CNN}}$. We hypothesise this input will allow S-CNNs to be robust to a change in gradient scheme at inference time as long as the diffusion sensitising factors are the same.

Another benefit of S-CNN's is the rotationally equivariant architecture \cite{Cohen2018SPHERICALCNNS}. We hypothesise that this property will allow S-CNNs to extrapolate from a training dataset with a common primary fibre orientation and, during the inference stage, well estimate tissue with fibres oriented along any direction. As a result, S-CNNs do not require a diverse set of tissue microstructure orientations in the training dataset, reducing demands on the training dataset.

%% file: 02method.tex
\section{Experiments}
\label{sec:method}
Each claim made in this paper is evaluated with an individual experiment. The first experiment evaluates network robustness to differing gradient schemes; the second assesses network robustness to the distribution of the primary fibre orientations in the training set.

\begin{figure}[ht!]
\begin{center}
\includegraphics[width=\linewidth] {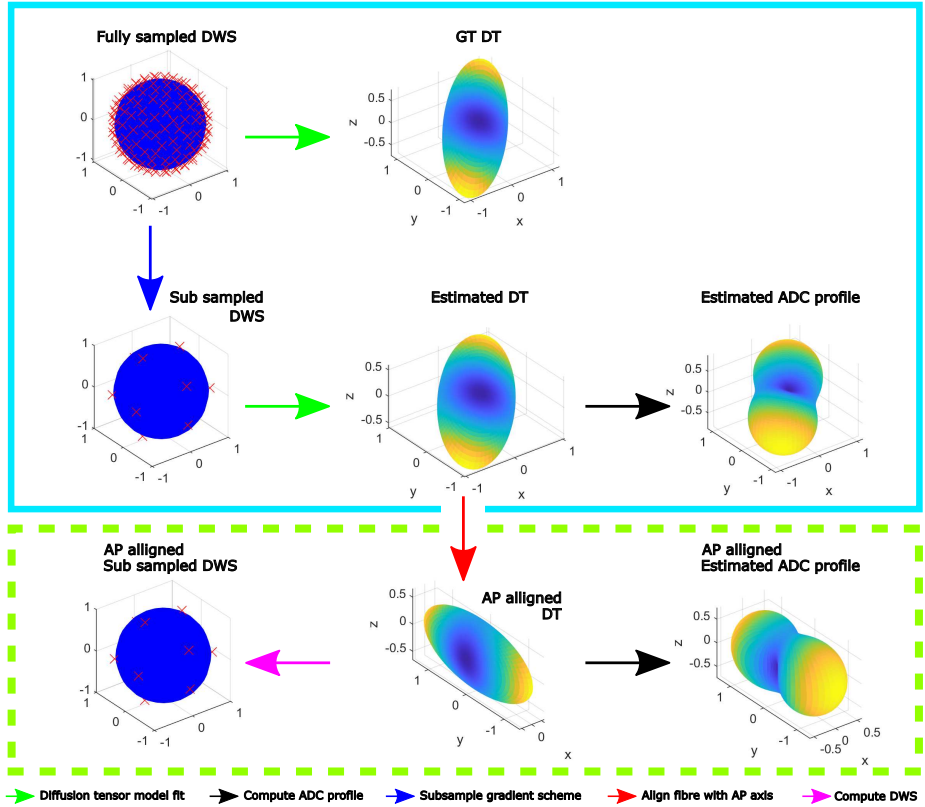}
\caption{
This figure shows how the testing data is generated for the both experiments (blue box). The figure also shows how the training data is generated for experiment 1 (blue box) and experiment 2 (green box). In the blue box, from the 90 directional $b=0$ normalised diffusion weighted signals (DWS) we estimate the ground truth diffusion tensor, and generate a 6-dir clinical scan by subsampling. The clinical scan's measurements are used as input to the FCN, but for compatibility with the S-CNN they must first be re-parameterised as an ADC profile. This is achieved by exploiting the 1-to-1 relationship described in section \ref{subsec:methods/GenDenseSampledSphSignals}. For the second experiment training data is required that has the primary fibre oriented along the AP axis. This is achieved by changing the orientation of the estimated DT and calculating the 6 DWS, for FCN, or calculating the ADC profile for S-CNN.
}
\label{fig:methods}
\end{center}
\end{figure}

\subsection{Experiment 1}
\subsubsection{Study design}
In this experiment we test if the networks are robust to new gradient schemes at inference time. In order to show this we propose an experiment where both networks are trained with a typical gradient scheme and then, in the inference phase, the trained networks are applied to data collected using a gradient scheme (1) the same as the training gradient scheme and another (2) different to the training scheme.

\subsubsection{Network architectures and training parameters}
\label{sec:method:subsec:Networks}
Both network architectures are voxel-wise networks. The S-CNN architecture we use, known as the hybrid spherical CNN architecture \cite{Cobb2021EfficientCNNs}, was chosen as it has been shown to be highly rotationally equivariant whilst also being computationally efficient. The specific network parameters follow the spherical MNIST experiment and the input to this network is a densely sampled spherical signal, described later. The FCN network architecture implementation, used as the baseline, is consistent with the established FCN techniques for dMRI parameter estimation \cite{Golkov2016Q-SpaceScans, Aliotta2019HighlyNetworks}. This network input follows the standard practice consisting of 6 $b=0$ normalised diffusion-weighted signals. The network hyperparameters are: three hidden layers with number of units=[100,100,10] and ReLU activation function.

The training parameters are chosen in order for FCN to perform optimally and consistent between the networks so that any difference between trained networks is solely because of their architectures. To achieve this, the training parameters are consistent with the FCN literature; specifically the training regime used Adam optimiser for 50 epochs with learning rate set to 0.001, the batch size 32 and the loss metric MSE.  

\subsubsection{Generating densely sampled spherical signals}
\label{subsec:methods/GenDenseSampledSphSignals}
S-CNNs require densely sampled spherical signals as input. Densely sampled spherical signals are generated for each voxel by utilising the property - described earlier in Section \ref{sec:dMRIparamEstim:subsec:DTexample} - of the 1-to-1 mapping between six-directional dMRI signals and the 6 independent values of the diffusion tensor. Due to this property, six-directional dMRI signals - with all of their noise - are perfectly and uniquely described by a DT. From this DT, $\Tilde{D}$, a spherical function called the ADC profile may be derived, following:
\begin{equation}
    ADC(\textbf{g};\Tilde{D}) = \textbf{g}^T \Tilde{D} \textbf{g},
    \label{egn:ADC}
\end{equation}
with gradient direction $\textbf{g}$. By densely sampling the gradient directions and calculating the resultant ADC, we generate the input required for S-CNNs. The process of generating densely sampled spherical signals is visually described in Figure \ref{fig:methods}.

\subsubsection{Datasets}
A dataset is required for training and testing the models. All of the deep learning models evaluated in this paper are supervised machine learning techniques, therefore, the training dataset must consist of a set of input values paired with ground truth output values. For the high-quality ground truth output, a dataset is required that contains a sufficiently large number of DWIs to provide accurate estimation of FA.

For this reason, we have chosen dMRI data from the Human Connectome Project (HCP) which includes 90 DWIs at b=1000 s/mm$^2$. Ground-truth FA maps are computed from the complete set of DWIs. Subsets of six-directional DWIs are sampled from the 90 DWIs to be in maximal agreement with the chosen gradient schemes. Training was performed with one participant and, to show generalisation, data from 12 unseen participants was used for testing.

\subsubsection{Evaluation}
Both networks are trained using diffusion weighted signals corresponding to one gradient scheme, the Skare scheme \cite{Skare2000ConditionMRI}, and evaluated with signals corresponding to both the original Skare scheme as well as a new scheme, Jones \cite{Jones1999OptimalImaging}. Quantitative measurements of the estimation error are calculated with root mean square error (RMSE) over specific FA ranges over the whole image. Statistical significance between distributions is quantified with paired t-tests. Qualitative assessments are made using maps of estimates and errors relative to the ground truth.

\subsection{Experiment 2}
\subsubsection{Study design}
In this experiment we test the network's robustness to the distribution of the primary fibre orientations in the training set by testing the networks on a set of microstructure configurations whose primary fibre orientation lies both inside and outside of the training dataset's distribution. We achieve this by restricting the primary fibre orientation in the training dataset to align with the anterior-posterior axis and testing on microstructure oriented in all directions. We hypothesise that networks robust to the distribution of primary fibre orientation in the training set will estimate the FA equally well independent of the direction of the primary fibre orientation.

\subsubsection{Restricting primary fibre orientation}
\label{subsec:methods/exp2}
The training distribution of the primary fibre orientations is restricted to the anterior-posterior axis by adapting the dense ADC sampling algorithm.
In the dense ADC sampling algorithm, after a noisy DT's estimation, the noisy DT's shape and size are extracted by eigendecomposition. Next, the primary fibre orientation, otherwise known as the principal eigenvector, is set to the anterior-posterior axis whilst the secondary eigenvector is set to the superior-inferior axis; see visual description in the light green dashed box in Figure \ref{fig:methods}. This new AP-aligned diffusion tensor is used to generate the input required for FCN, by computing the 6 directional $b=0$ normalised diffusion-weighted signals from the DT forward model, and the input required for S-CNN, by densely sampling the ADC profile.

\subsubsection{Evaluation}
Network architectures, training scheme, training dataset and evaluation metrics are the same as experiment 1. Only one gradient scheme is required for this experiment so the Skare scheme is chosen for training and testing. Training data undergoes primary fibre orientation restriction, whilst testing data is unrestricted. To show further benefits of rotational equivariance we test to see if this property allows S-CNNs to estimate with high fidelity when starved of training data points. For this, an S-CNN is trained with only 10 \% of the total training datapoints. The distribution of estimation error over the primary fibre orientation is evaluated.

%% file: 03resultsNdiscussion.tex
\section{Results and Discussion}
\label{sec:resultsNdiscussion}

\subsection{For Experiment 1}
\begin{figure}[ht!]
\begin{center}
\includegraphics[width=\linewidth] {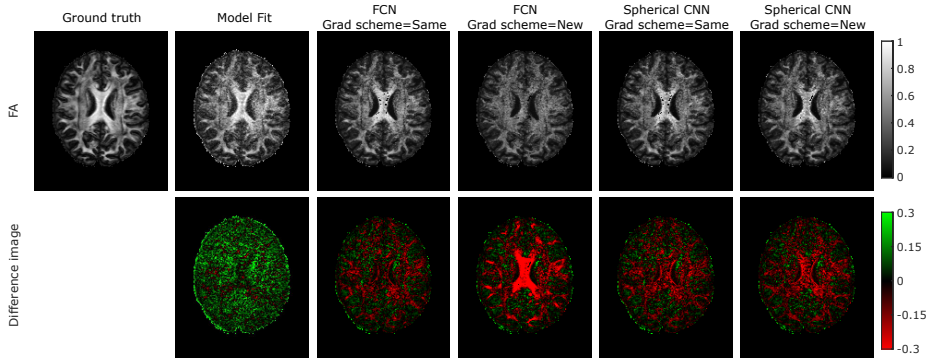}
\caption{
Results of experiment 1 are visualised on an subject unseen during training. Conventional fully-connected networks are compared against the S-CNN model at estimating FA with either the same gradient scheme as used in training or a new gradient scheme. The FCN shows a drop in performance when the gradient scheme is different between training and testing. The S-CNN doesn't have this issue, therefore, it is robust to differing gradient schemes.
}
\label{fig:experiment1}
\end{center}
\end{figure}

\begin{figure}[ht!]
\begin{center}
\includegraphics[width=\textwidth] {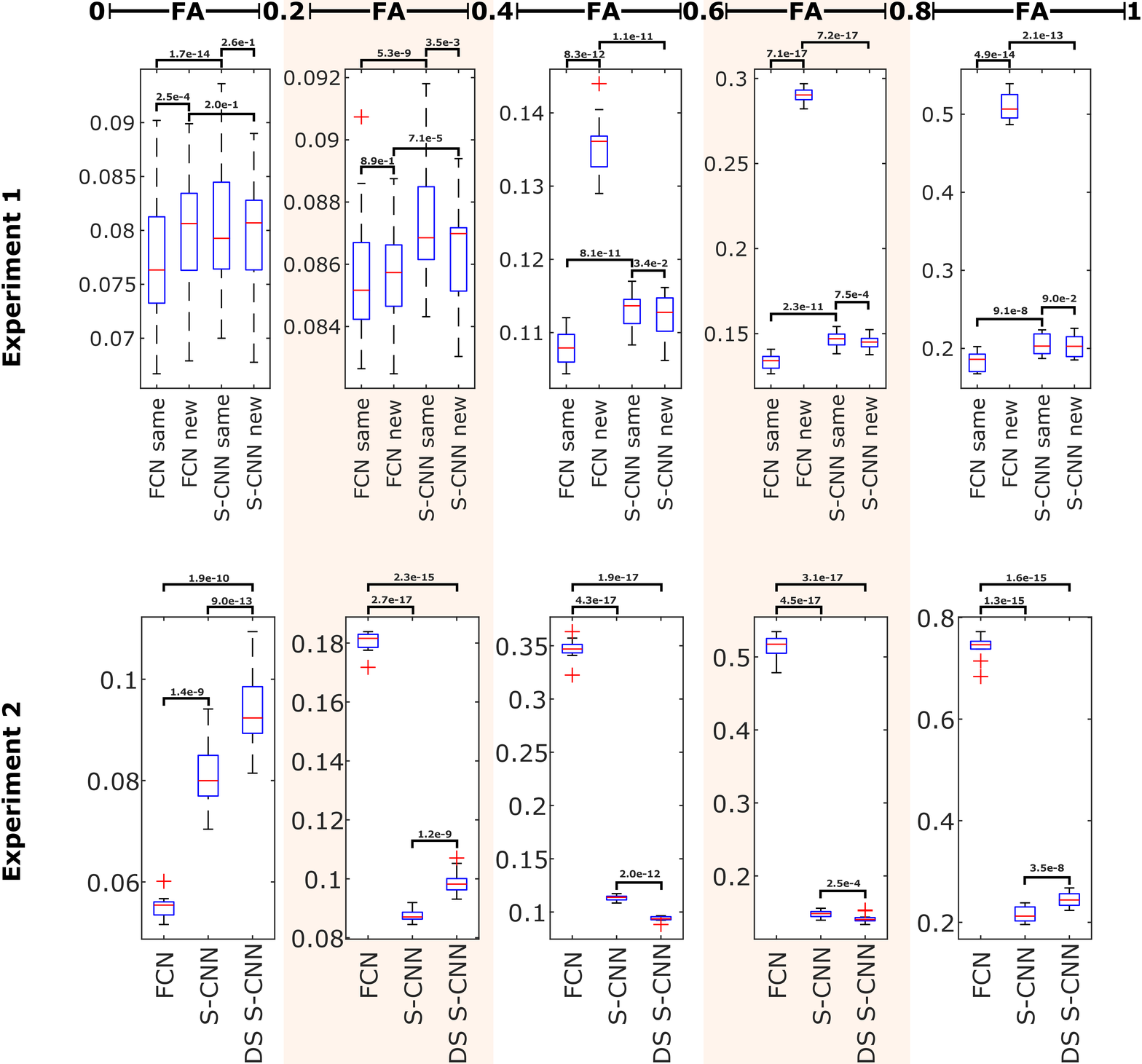}
\caption{
Boxplots of the RMSE over the 12 testing subjects for both experiments, with p-values between distributions from paired t-test shown. The testing data is uniformly split into 5 GT FA ranges. For both experiments the problems with FCN become apparent when applied to anisotropic signal. 
}
\label{fig:boxplots}
\end{center}
\end{figure}

Qualitative and quantitative results of experiment 1 are shown in figures \ref{fig:experiment1} and \ref{fig:boxplots} respectively.

Figure \ref{fig:experiment1} shows an example slice of a GT FA map from a test subject along with the model fitting, FCN and S-CNN estimations and error maps when the gradient scheme is the same or different between training and testing. When the training and testing schemes are the same FCN's performance is consistent with the literature, estimating FA faithfully. When the gradient schemes are different the FCN estimates poorly. Estimation is especially poor in areas of high FA, such as the corpus callosum. S-CNN's estimations are similar to the GT regardless of the gradient scheme. This shows S-CNN's robustness to differing gradient schemes.


\begin{figure}[hb!]
\begin{center}
\includegraphics[width=\linewidth] {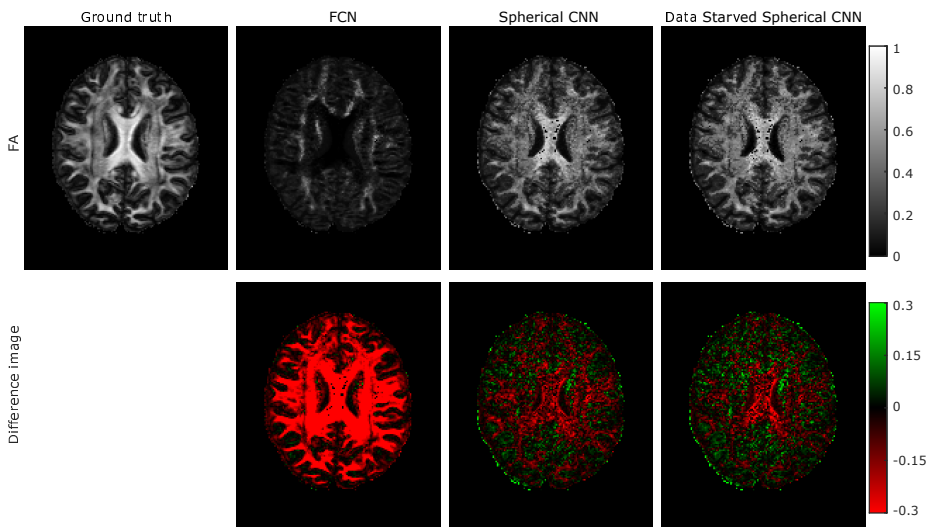}
\caption{
Comparison between the FCN model and the S-CNN model at estimating the FA when trained only with microstrucure oriented along the anterior-posterior axis. These are also compared against a training data starved S-CNN. We see that spherical CNN outperforms FCN as the noise is greatly reduced and less structured. Data starved S-CNN performs similarly well to the S-CNN trained using the full dataset.
}
\label{fig:experiment2a}
\end{center}
\end{figure}

\begin{figure}[ht!]
\begin{center}
\includegraphics[width=\linewidth] {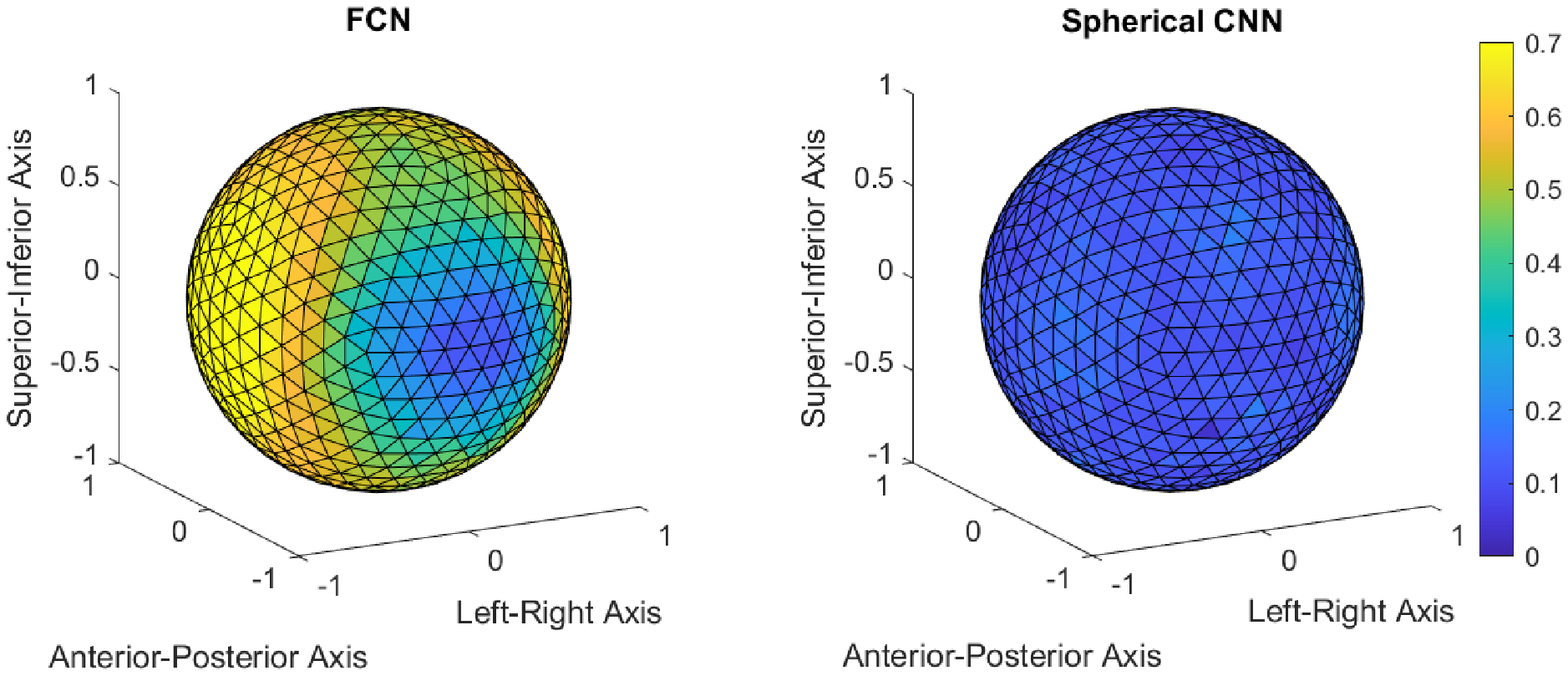}
\caption{
The distribution of the estimation error over the sphere is compared between the fully-connected network and spherical CNN. To colour each tile we find the voxels whose GT primary fibre orientation lies within the tile's surface and compute the mean RMSE. For this evaluation only voxels with FA $>$ 0.6 were used. We see that FCN's error is only low along the AP axis and high in all other directions. S-CNN's error is low over the entire surface of the sphere.
}
\label{fig:experiment2bSpherical}
\end{center}
\end{figure}
Figure \ref{fig:boxplots} reinforces the qualitative observations with quantitative measures showing boxplots of the mean RMSE over the 12 subjects. 
We see for anisotropic signals (FA $>$ 0.4) the conventional FCN performance is significantly worse (p $<$ 8e-12) when applied to a new gradient scheme than on the training scheme, whereas, the S-CNN models estimation fidelity does not decrease when applied to a new scheme.
In both figures, on differing gradient schemes the FCN model is shown to be more inaccurate in regions of high FA. 
This loss of accuracy in high FA regions may be caused by the signal attenuation in these regions being greatly dependant on the direction of measurement. Therefore, for low FA voxels the DWS from any two gradient schemes are similar. Whereas, for areas of high FA the DWS will be vastly different and therefore the training distribution will be different to the testing distribution.

Gradient scheme robustness may also be achieved with FCN by first encoding the signals in terms of their corresponding spherical harmonic representation \cite{Lin2019FastNetwork}. This approach is essentially equivalent to S-CNN but does not offer the property of rotational equivariance, the effect of which is demonstrated in the next subsection.

\subsection{For Experiment 2}
The results of experiment 2 are shown in figures \ref{fig:boxplots}, \ref{fig:experiment2a} and \ref{fig:experiment2bSpherical}. Figure \ref{fig:experiment2a} qualitatively shows the effect of estimating the full brain volume using networks trained only on tissue microstructure aligned with the anterior-posterior axis. 
The FCN model consistently underestimates FA in regions where the underlying tissue microstructure does not align with anterior-posterior direction (e.g. the corpus callosum which consists of left-right white matter tracts). 
In contrast, the S-CNN models accurately estimate FA independently of the primary fibre direction, and the error is far less structured than the FCN's.

This is mirrored in the quantitative measurements over the 12 testing subjects, shown in figure \ref{fig:boxplots}. 
The FCN model only well estimates isotropic signals (FA$<$ 0.2); however the performance difference between different methods is relatively small, with the difference between the mean RMSE for different methods no larger than 0.04. 
When the signal becomes anisotropic, the FCN model estimates the signal significantly worse than the S-CNN models, with the difference in the mean RMSE for different methods no smaller than 0.08.
As the signals get more anisotropic the FCN model performs even worse until in the top FA bracket (0.8$\leq$FA$<$1) the mean RMSE between the FCN and the S-CNN models is 0.5, 10 times the difference in error of the isotropic signals.

Figure \ref{fig:experiment2bSpherical} shows the distribution of the absolute error over the full range of primary fibre orientations for anisotropic signals. The FCN model well estimates tissue microstructure aligned with the anterior-posterior axis, seen during training. However the error quickly grows as the primary fibre orientation deviates from this axis. This adverse feature is not exhibited by the S-CNN model as the estimation error is low and independent of training dataset distribution of the primary fibre orientation. The lack of rotational equivariance in FCNs hinders estimation performance when generalising to microstructure with primary fibre orientation not sampled in the training distribution. This has potential for orientation bias in the training dataset to lead to poor estimation for under-sampled directions.

In figures \ref{fig:boxplots} and \ref{fig:experiment2a} another advantage of rotational equivariance is shown. Performance of the S-CNN network is not greatly changed when only a tenth of the training dataset is used. This is shown in the maps of FA estimated by the data starved S-CNN model in figure \ref{fig:experiment2a} and the RMSE over the 12 testing subjects shown in figure \ref{fig:boxplots}. The robustness to a lack of training data is due to S-CNNs being rotationally equivariant. Therefore, during training only a diverse distribution of the microstructure shape is required and not their orientations as well. This property of S-CNNs is a real benefit as it reduces the number of training datapoints required for good estimation at inference stage and, when simulating training data, allows for denser sampling of the shape parameters as SO(3) need not be sampled.

%% file: 04conclusion.tex
\section{Conclusion}
\label{sec:conclusion}
In this work we explore the advantages of S-CNNs for dMRI parameter estimation over conventional FCNs. Representing diffusion weighted signals as a spherical image is here demonstrated to gain robustness to the gradient scheme absent from conventional FCNs, at no cost to fidelity. This removes the need to retrain a new network for every gradient scheme, a feature especially beneficial when combining data from multiple sites. S-CNN is shown to be superior to FCN methods additionally because of its rotational equivariance property. This enables the network to encode information about the pattern of the signal irrespective of primary fibre orientation. This eliminates the need to sample diffusion primary fibre orientations, reducing the number of samples needed to cover the full parameter space.

%% file: main.bbl
\begin{thebibliography}{10}
\providecommand{\url}[1]{\texttt{#1}}
\providecommand{\urlprefix}{URL }
\providecommand{\doi}[1]{https://doi.org/#1}

\bibitem{Aliotta2019HighlyNetworks}
Aliotta, E., Nourzadeh, H., Sanders, J., Muller, D., Ennis, D.B.: {Highly
  accelerated, model-free diffusion tensor MRI reconstruction using neural
  networks}. Medical Physics  \textbf{46}(4),  1581--1591 (5 2019).
  \doi{10.1002/mp.13400},
  \url{https://aapm.onlinelibrary.wiley.com/doi/abs/10.1002/mp.13400}

\bibitem{Basser1994EstimationEcho}
Basser, P.J., Mattiello, J., LeBihan, D.: {Estimation of the Effective
  Self-Diffusion Tensor from the NMR Spin Echo}. Journal of Magnetic Resonance,
  Series B  \textbf{103}(3),  247--254 (3 1994). \doi{10.1006/jmrb.1994.1037}

\bibitem{Bodini2014DiffusionDisorders}
Bodini, B., Ciccarelli, O.: {Diffusion MRI in Neurological Disorders}.
  Diffusion MRI: From Quantitative Measurement to In vivo Neuroanatomy: Second
  Edition pp. 241--255 (1 2014). \doi{10.1016/B978-0-12-396460-1.00011-1}

\bibitem{Chen2020EstimatingNetworks}
Chen, G., Hong, Y., Zhang, Y., Kim, J., Huynh, K.M., Ma, J., Lin, W., Shen, D.,
  Yap, P.T.: {Estimating Tissue Microstructure with Undersampled Diffusion Data
  via Graph Convolutional Neural Networks}. In: MICCAI 2020. vol. 12267 LNCS,
  pp. 280--290. Springer Science and Business Media Deutschland GmbH (2020).
  \doi{10.1007/978-3-030-59728-3{\_}28}

\bibitem{Cobb2021EfficientCNNs}
Cobb, O.J., Wallis, C.G.R., Mavor-Parker, A.N., Marignier, A., Price, M.A.,
  D'Avezac, M., McEwen, J.D.: {Efficient Generalized Spherical CNNs}. In: ICLR
  2021 (2021), \url{http://arxiv.org/abs/2010.11661}

\bibitem{Cohen2018SPHERICALCNNS}
Cohen, T.S., Geiger, M., K{\"{o}}hler, J., Welling, M.: {Spherical CNNs}. In:
  ICLR 2018 (2018), \url{http://arxiv.org/abs/1801.10130}

\bibitem{Elaldi2021EquivariantData}
Elaldi, A., Dey, N., Kim, H., Gerig, G.: {Equivariant Spherical Deconvolution:
  Learning Sparse Orientation Distribution Functions from Spherical Data}. In:
  Lecture Notes in Computer Science (including subseries Lecture Notes in
  Artificial Intelligence and Lecture Notes in Bioinformatics). vol. 12729
  LNCS, pp. 267--278 (2021). \doi{10.1007/978-3-030-78191-0{\_}21}

\bibitem{Golkov2016Q-SpaceScans}
Golkov, V., Dosovitskiy, A., Sperl, J.I., Menzel, M.I., Czisch, M.,
  S{\"{a}}mann, P., Brox, T., Cremers, D.: {q-Space Deep Learning: Twelve-Fold
  Shorter and Model-Free Diffusion MRI Scans}. IEEE Transactions on Medical
  Imaging  \textbf{35}(5),  1344--1351 (2016). \doi{10.1109/TMI.2016.2551324}

\bibitem{Jones1999OptimalImaging}
Jones, D.K., Horsfield, M.A., Simmons, A.: {Optimal strategies for measuring
  diffusion in anisotropic systems by magnetic resonance imaging}. Magnetic
  Resonance in Medicine  \textbf{42}(3),  515--525 (1999).
  \doi{10.1002/(SICI)1522-2594(199909)42:3<515::AID-MRM14>3.0.CO;2-Q}

\bibitem{Jones2013GaussianSignal}
Jones, D.K.: {Gaussian Modeling of the Diffusion Signal}. In: Diffusion MRI:
  From Quantitative Measurement to In vivo Neuroanatomy: Second Edition, pp.
  87--104. Elsevier Inc. (11 2013). \doi{10.1016/B978-0-12-396460-1.00005-6}

\bibitem{Lin2019FastNetwork}
Lin, Z., Gong, T., Wang, K., Li, Z., He, H., Tong, Q., Yu, F., Zhong, J.: {Fast
  learning of fiber orientation distribution function for MR tractography using
  convolutional neural network}. Medical Physics  \textbf{46}(7),  mp.13555 (5
  2019). \doi{10.1002/mp.13555},
  \url{https://onlinelibrary.wiley.com/doi/abs/10.1002/mp.13555}

\bibitem{Park2021DIFFnet:B-values}
Park, J., Jung, W., Choi, E.J., Oh, S.H., Jang, J., Shin, D., An, H., Lee, J.:
  {DIFFnet: Diffusion parameter mapping network generalized for input diffusion
  gradient schemes and b-values}. IEEE Transactions on Medical Imaging  (2021).
  \doi{10.1109/TMI.2021.3116298}

\bibitem{PeterJ.Basser1996MicrostructuralMRI}
{Peter J.Basser}, {Carlo Pierpaoli}: {Microstructural and physiological
  features of tissues elucidated by quantitative-diffusion-tensor MRI}. Journal
  of Magnetic Resonance pp. 209--219 (1996)

\bibitem{Sedlar2021ADMRI}
Sedlar, S., Alimi, A., Papadopoulo, T., Deriche, R., Deslauriers-Gauthier, S.:
  {A Spherical Convolutional Neural Network for White Matter Structure Imaging
  via dMRI}. In: Lecture Notes in Computer Science (including subseries Lecture
  Notes in Artificial Intelligence and Lecture Notes in Bioinformatics). vol.
  12903 LNCS, pp. 529--539. Springer, Cham (9 2021).
  \doi{10.1007/978-3-030-87199-4{\_}50},
  \url{https://link.springer.com/chapter/10.1007/978-3-030-87199-4_50}

\bibitem{Skare2000ConditionMRI}
Skare, S., Hedehus, M., Moseley, M.E., Li, T.Q.: {Condition Number as a Measure
  of Noise Performance of Diffusion Tensor Data Acquisition Schemes with MRI}.
  Journal of Magnetic Resonance  \textbf{147}(2),  340--352 (2000).
  \doi{10.1006/jmre.2000.2209}, \url{https://pubmed.ncbi.nlm.nih.gov/11097823/}

\bibitem{Zhang2012NODDI:Brain}
Zhang, H., Schneider, T., Wheeler-Kingshott, C.A., Alexander, D.C.: {NODDI:
  practical in vivo neurite orientation dispersion and density imaging of the
  human brain}. NeuroImage  \textbf{61}(4),  1000--1016 (2012).
  \doi{10.1016/j.neuroimage.2012.03.072},
  \url{http://www.ncbi.nlm.nih.gov/pubmed/22484410}

\end{thebibliography}
